\documentclass{elsart}
\usepackage{natbib}
\usepackage{graphicx}
\usepackage{dcolumn}
\usepackage{bm}
\usepackage{amsmath}
\usepackage{amsfonts}
\usepackage{amssymb}
\newcommand{\ket}[1]{|#1\rangle}

\newcommand{\braket}[3]{\langle#1|#2|#3\rangle}

\begin{document}

\begin{frontmatter}
\title{Markovian Behaviour and Constrained Maximization of the Entropy in Chaotic Quantum Systems}
\author[IFFI]{A. Romanelli}
\author[IFFI]{A.C. Sicardi Schifino\thanksref{IFFC}}
\author[IFFI]{G. Abal}
\author[IFFI]{R. Siri}
\author[UFRJ]{R. Donangelo}
\address[IFFI]{Instituto de F\'{\i}sica, Facultad de Ingenier\'{\i}a\\
Universidad de la Rep\'ublica\\ C.C. 30, C.P. 11000, Montevideo, Uruguay}
\address[UFRJ]{Instituto de F\'{\i}sica, Universidade Federal do Rio de Janeiro\\ 
C.P. 68528, 21945-970 Rio de Janeiro,Brazil}
\thanks[IFFC]{also at: Facultad de Ciencias, Universidad de la Rep\'ublica.}
\date{\today}
\begin{abstract}
The separation of the Schr\"{o}dinger equation into a Markovian and an interference term provides a new insight 
in the quantum dynamics of classically chaotic systems. The competition 
between these two terms determines the localized or diffusive character of the dynamics. 
In the case of the Kicked Rotor, we show how the constrained maximization of the entropy implies 
exponential localization. 
\end{abstract}
\end{frontmatter}

In the last few decades the field of Quantum Chaos has drawn the attention of
researchers in several areas of science. Many interesting phenomena were studied
in systems such as atom traps and microwave cavities. Furthermore, the recent
advances in technology that allow to construct and almost perfectly preserve 
quantum states, has opened the field of quantum computation, where chaotic
effects and their control play an essential role \cite{QC}. 

In this work we develop a different and general approach to the subject of
dynamical localization (DL) and the related issue of quantum diffusion. This
approach leads to an improved understanding of why DL takes place in some
systems while in others quantum diffusion continues for ever. The path we
follow consists in rewriting the Schr\"{o}dinger equation in a form in which
the part responsible for DL is separated from the part responsible for quantum
diffusion. As we shall see, the terms responsible for quantum diffusion have
the form of a master equation, typical of Markovian processes, while the other
part consists of interference terms required to preserve the unitary character
of the quantum evolution. The master equation has been extensively used by W.
Zurek and co--workers \cite{Zurek} to study the emergence of a classical
behavior due to environment--induced decoherence. For the systems considered
in the present work the environment does not play a role, and the decoherence,
if present,  is self--generated by the dynamics.

Consider a quantum system described by a generic time-dependent Hamiltonian of
the form $H(t)~=~H_{0}~+~V(t)$, where $H_{0}$ is the time-independent part,
with known eigenstates and eigenvalues satisfying $H_{0}\ket{k}=E_k\ket{k}$. We
write the wave function as $\ket{\Psi (t)}=\sum_{k}a_k(t)\ket{k}$. The
time-dependent part of the Hamiltonian, $V(t)$, induces transitions between the
eigenstates of $H_{0}$. The actual quantum unitary evolution of $\ket{\Psi}$ is
given by the Schr\"{o}dinger equation. However, if discrete times $t_n$ are 
considered, then a quantum map is obtained

\begin{equation}
a_{k}(t_{n+1})=\sum_{l}U_{kl}^{(n)}a_{l}(t_{n}),
\label{mapa1}
\end{equation}
where $U_{kl}^{(n)}\equiv\braket{k}{U(t_{n+1},t_{n})}{l}$ and
$U(t_{n+1},t_{n})$ is the evolution operator connecting the state at time
$t_{n}$ with the one at time $t_{n+1}$. In terms of $U$, the occupation probability
$P_{k}(t_{n})\equiv \left|a_k(t_n)\right|^2$ can be expressed as

\begin{equation}
P_{k}(t_{n+1})=\sum_{l,m}U_{kl}^{(n)}U_{km}^{(n)}{}^\ast
a_{l}(t_{n})a_{m}^{\ast}(t_{n}).
\label{master1}
\end{equation}

If we separate the diagonal terms in its rhs, eq.~(\ref{master1})
may be written as

\begin{equation}
P_{k}(t_{n+1})=\sum_{l}T_{kl}P_{l}(t_{n})+\beta_k(t_n)\,,
\label{master2}
\end{equation}
with

\begin{equation}
\beta_{k}(t_n)=\sum_{\stackrel{\scriptstyle l,m}{(l\neq m)}}
U_{kl}^{(n)}U_{km}^{(n)}{}^{\ast}a_{l}(t_{n})a_{m}^{\ast}(t_{n})\,,
\label{beta}
\end{equation}
where we have defined $T_{kl}\equiv\left| U_{kl}^{(n)}\right| ^{2}$ as the
transition probability. This is meaningful because $T_{kl}\geq 0$ and
$\sum_{k}T_{kl}=\sum_{l}T_{kl}=1$. Thus, if the term $\beta_k$
could be neglected in eq.~(\ref{master2}), the time evolution of the occupation
probability would be described by a Markovian process in which the transition
probability $k\rightarrow l$, in a time $\Delta t_{n}=t_{n+1}-t_{n},$ is given
by $T_{kl}$.

We now consider the transition probability per unit time, defined as

\begin{equation}
W_{kl}\equiv\frac{T_{kl}-\delta_{kl}}{\Delta t_{n}},
\label{Wkl}
\end{equation}
where $\delta_{kl}$ is the Kronecker delta. A straightforward calculation shows
that the quantum evolution equation~(\ref{master2}), can be written as

\begin{equation}
P_{k}(t_{n+1})=P_{k}(t_{n})+
\sum_{l\neq k} \left( W_{kl}P_{l} - W_{lk}P_{k}\right)
\Delta t_{n}+\beta_{k}.
\label{master3}
\end{equation}

In this form, two qualitatively different terms can be distinguished: one
associated with a Markovian process, {\it i.e.} a classical-like diffusion, and
the other, $\beta_{k}$, associated to quantum interference effects. This last
term preserves the unitary character of the evolution. Its contribution depends
on the continuous or discrete character of the dynamical response spectrum
\cite{jain,peres,2frecuencias}. For systems with a discrete spectrum, as the
periodic kicked rotor, it is of the same order of magnitude  as the Markovian
term, and becomes responsible for the dynamical localization found in such
systems. A finite time of the order of $1/\Delta\omega$, where $\Delta\omega$
is the average separation in the frequency response spectrum, is required in
order to resolve the discreteness of the spectrum. At shorter time scales, the
discreteness of the spectrum has no effect on the dynamics, the Markovian
approximation holds and the system ``mimics'' classical chaos. In the case of a
continuous spectrum, the terms contributing  to $\beta_k$ in eq.~(\ref{beta})
are scattered on the complex plane in such a way that their sum  becomes
negligible compared to the Markovian term in eq.~(\ref{master3}). Then, the 
unitary evolution is well approximated, for arbitrary long times, by a
Master equation 
\begin{equation}
\frac{\partial P_{k}}{\partial t_n} = 
\sum_{l\neq k} \left( W_{kl}P_{l} - W_{lk}P_{k}\right)
\label{maestra}
\end{equation}
which results from eq.~(\ref{master3}) if $\beta_k$ is neglected. 

Assuming, for simplicity, that  $W_{k,k-l}=W_{k,k+l}$, eq.~(\ref{maestra}) may
be written as a diffusion equation,
\begin{equation}
\frac{\partial P_{k}}{\partial t_n} = \frac{D}{2}\frac{\partial^2 P_k}
{\partial k^2}
\label{master4}
\end{equation}
where the diffusion coefficient is
\begin{equation}
D = 2\sum\limits_{l=1}^{\infty}W_{k,k+l}l^2.
\label{D_general}
\end{equation}
The differential operators in eqs.~(\ref{maestra}) and (\ref{master4}) represent
the discrete derivatives
$\partial P_k/\partial t_n~=~\left[ P_k(t_{n+1})-P_k(t_n)\right]/\Delta t_n$ and
$\partial^2 P_k/\partial k^2~=~\left[P_{k+l}(t_{n})+P_{k-l}(t_n)-2P_k(t_n)\right]/l^2$.
Note that, according to eq.~(\ref{master4}), an initial gaussian distribution evolves as a spreading gaussian.

As it is well known \cite{vankampen}, a Markovian process described by a Master
equation such as (\ref{maestra}), satisfies the H-theorem, which assures that
the thermodynamic entropy, $S\equiv-\sum_{k}P_{k}\ln P_{k}$ can only increase.
On the other hand, in a unitary evolution the entropy calculated from the
density operator, {\bf S}$\equiv$~-~trace$(\rho\ln \rho )$, must be constant.
We note that this expression is equivalent to the entropy $S$ only if the
index $k$ can be associated to the eigenstates of the complete Hamiltonian
$H(t)$. Obviously in this situation the Master equation does not hold. Since
in our case we have used the representation of eigenstates of $H_{0}$, and not
of $H(t)$, the thermodynamical entropy $S$ corresponds to a ``coarse-graining''
of the entropy {\bf S}. In the rest of this paper, we refer to the
thermodynamic entropy $S$ simply as ``entropy''. Even though it is possible
to define quantum dynamical entropies \cite{Alicki}, as long as the Master
equation (\ref{maestra}) holds it is adequate to use the associated 
thermodynamical entropy $S$. 

The introduction of the entropy $S$ will allow us to
use thermodynamic arguments to draw interesting conclusions regarding DL.
We recall \cite{Haken} that if the value of an observable $M$ in state
$\ket {k}$ is $M_{k}$, the maximum value of the entropy consistent with a given
constraint,
\begin{equation}
\sum _k M_{k}P_{k}=\langle M\rangle,
\label{constraint}
\end{equation}
corresponds to an exponential (canonical) distribution
\begin{equation}
  P_{k}~=~e^{-\Omega}e^{-\lambda M_{k}}. 
\label{canonica}
\end{equation}
This equilibrium distribution is attained after a diffusive process in
which the entropy is maximized. The Lagrange multipliers $\lambda,\Omega$ are
determined by the constraint (\ref{constraint}) and the normalization condition
$\sum _k P_{k}=1$, respectively.  In particular, the condition
$\frac{\partial\Omega}{\partial\lambda}+\langle M\rangle =0$ must be satisfied.
However, if the constraint  (\ref{constraint}) is removed, the distribution
evolves according to a diffusion equation for arbitrary long times and the
entropy increases without bound, provided that the spectrum of $H_0$ is unbounded.

At this point, it is convenient to consider a concrete example.
We choose the quantum Kicked Rotor (QKR), one of the first classically chaotic
systems to be quantum-mechanically investigated \cite{Casati79}.
For this system, the evolution operator has the following form \cite{Izrailev}

\begin{equation}
U_{kl}^{(n)}=i^{-(l-k)}J_{l-k}(\kappa)  e^{-iE_{l}\Delta t_{n}/\hbar} ,
\label{QKR_evop}
\end{equation}
where $\kappa=K/\hbar$ is the dimensionless kick strength, $E_{l}~=~\hbar^2l^2/2I$
is the kinetic energy of the rotor, $I$ its moment of inertia,
$\Delta t_{n}=t_{n+1}-t_{n}=T$ the constant time interval between kicks and $J_s$
is the $s^{th}$ order Bessel function. Note that in this case, the condition
$W_{k,k-l}=W_{k,k+l}$ is satisfied by (\ref{QKR_evop}).
Then, as long as the term $\beta_k$ can be neglected in eq.~(\ref{master3}),
the evolution is given by the diffusion equation (\ref{master4}).
The diffusion coefficient, obtained from eq.~(\ref{D_general}) using standard
properties of the Bessel functions, is $D=\kappa^2/2T$ , which is consistent
with the classical result.

\begin{figure}
\centerline{\includegraphics[scale=0.6, angle=-90]{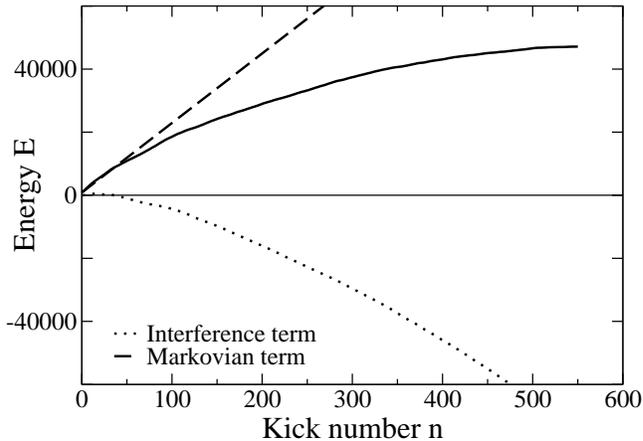}}
\caption{\label{fig:energy}
\footnotesize 
Contribution of each of the terms of eq.~(\ref{delta E}) to the average energy
of the kicked rotor. See text for details.}
\end{figure}

The average energy of the QKR is given by $E(t_{n})~=~\sum_{k=-\infty}^{\infty
}E_{k}P_{k}(t_n)$. A direct calculation based on eq.~(\ref{master3}),  shows
that the energy change across two consecutive kicks, $\Delta E_n\equiv
E(t_{n+1})-E(t_n)$, can be written in terms of the transition probabilities
$W_{kl}\Delta t_n $ as 

\begin{eqnarray} 
\Delta E_n&=&
\sum_{k,l=-\infty}^{\infty}E_{l}P_{k}(t_{n})W_{kl}\Delta t_n
+\sum_{l=-\infty}^{\infty}E_{l}\beta_{l}\nonumber\\ &=&\frac{K^2}{4I}
+\sum_{l=-\infty}^{\infty}E_{l}\beta_{l}(t_{n}). 
\label{delta E} 
\end{eqnarray}
The first term in the rhs of this expression describes the classical diffusive
increase in the energy while the second term, involving the sum  of $\beta_l$,
accounts for quantum interference effects and is calculated explicitly in
\cite{2frecuencias}. In Fig.~\ref{fig:energy}, we show the contribution of each
term in eq.~(\ref{delta E}) to the energy of the QKR, as a function of the
number of kicks, $n$. The dotted line in this figure corresponds to the
contribution of the interference term,  $\sum_l E_{l}\beta_{l}(t_{n})$ and the
dashed line to the contribution of the Markovian term. This last contribution 
has a slope $\kappa^2/2$ and is coincident with the classical linear diffusive
increase. After $\sim 50$ kicks, the cumulative effect of the quantum coherence
begins to cancel out the diffusive growth of the energy. In eq.~(\ref{delta
E}), $\Delta t_n = T$ and the parameters are $\kappa = 21.0$ and $\hbar T/I=1$.
The energy is in units of $\hbar^2/2I$ and the smoothing was obtained by
averaging over one hundred initial conditions, corresponding to the fifty lowest
eigenvalues of $H_0$. We see that the contribution of the interference term is
negligible until dynamical localization sets in, so that this term may be
identified as the one responsible for DL.

\begin{figure}
\centerline{\includegraphics[scale=0.6, angle=-90]{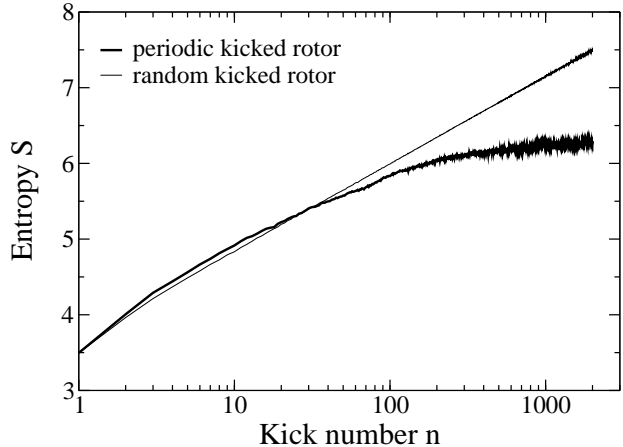}}
\caption{\label{fig:entropy}
\footnotesize Entropy $S=-\sum_k P_k {\rm ln} P_k$ as a function of the
kick number (log scale) for the quantum kicked rotor with a fixed time interval
between kicks (thick line) and a random time interval between kicks (thin line).
The parameters and averaging over initial conditions are the same as in
Fig.~\ref{fig:energy}.}
\end{figure}

During the quantum diffusion which precedes DL, when the interference term is
negligible, the quantum evolution mimics a Markovian process. 
 and the entropy
increases, as shown in Fig.~\ref{fig:entropy}. However, the discrete
quasi--energy spectrum of the QKR Floquet operator (\ref{QKR_evop}) introduces
a constraint that limits the entropy increase. In fact, as is well known
\cite{Izrailev}, only a finite number ($L_{0}$ on the average) of eigenstates
of $H_{0}$ are required to describe an eigenstate of the Floquet operator. If
we initialize the system in the  eigenstate $\ket{k_0}$ of $H_{0}$, the
subsequent evolution can only involve eigenstates $\ket{k}$ of $H_{0}$ such
that $\left| k-k_{0}\right|\le L$. This fact represents a constraint on the
entropy increase, of the type given  by eq.~(\ref{constraint}). The actual
value of $L$ depends on the initial condition, but it must obey  $\left\langle
L\right\rangle =L_{0}$, where the average is over different initial states
$\ket{k_0}$.  For each $k_0$, we take $L$, {\it i.e.} the maximum distance
$\left| k-k_{0}\right|$ in angular momentum space present in the dynamical
evolution, as the observable $M$ of eq.~(\ref{constraint}). Then, the
corresponding constraint is $\langle L\rangle =\langle \left|
k-k_{0}\right|_{max} \rangle =L_{0}$.  From this constraint and the
normalization condition for the occupation probability, we obtain the Lagrange
multipliers $\lambda =\sinh ^{-1}(1/L_0)$,  $\Omega =\ln\left[
L_0+(L_0^2+1)^{1/2}\right]$. Then, the probability distribution (\ref{canonica})
resulting from the constrained maximization of the entropy is

\begin{equation} P_{k}=\frac{e^{-\lambda
\left|k-k_0\right|}}{L_0+(L_0^2+1)^{1/2}}
\approx \frac {1}{2L_0} e^{-\frac{\left|k-k_0\right|}{L_0}}
\label{canonica2}
\end{equation}
because $L_{0}\gg 1$. This exponential profile for the localized wave function
has been verified both numerically \cite{Izrailev} and experimentally
\cite{experimental} for the case of the QKR. When the maximum value of the
entropy consistent with this constrain is attained, the interference term in
the rhs of eq.~(\ref{master3}) becomes non-negligible and so the master equation
approximation does not hold. Then the entropy remains constant (see
Fig.~\ref{fig:entropy}).

The origin of DL may thus be traced to the discrete nature of the
dynamical response frequency spectrum (or equivalently for the QKR, the
quasi--energy spectrum), which determines the degree of randomness of the
$\beta _k$ values in the complex plane. This is discussed in further
detail in  \cite{2frecuencias}.

To clarify the relation of DL to the behavior of the second term in the
rhs of eq.~(\ref{delta E}) we now consider a simple modification of the QKR, in
which the time intervals $\Delta t_{n}$, are randomly chosen. In this case
eqs.~(\ref{QKR_evop}) and (\ref{delta E}) are valid, but the contribution of
the interference term in the rhs of eq.~(\ref{master3}) is negligible for
arbitrarily long times since the dynamical response frequency spectrum is
continuous and this system does not show DL \cite{2frecuencias}. In this case,
the evolution is always Markovian and it can be described by the diffusion equation
(\ref{master4}) with the diffusion coefficient given by eq.~(\ref{D_general}).

One might think that these diffusive dynamics is an artifact resulting from the
randomness of the time intervals between successive kicks.  That this is not
the case may be seen by considering the double Kicked Rotor (QDKR), a
quasiperiodic version of the QKR. The QDKR is obtained when a second series of
pulses is applied to the rotor, with an irrational ratio between the periods of
both trains of pulses. The evolution operator (\ref{QKR_evop}) also holds for
the QDKR, but the time intervals $\Delta t_{n}$ now form a pseudo-random
sequence. The response spectrum of this system is dense and it undergoes
unlimited quantum diffusion at a rate consistent with the classical diffusion
rate \cite{2frecuencias}. This implies that the contribution of the
interference term in eq.~(\ref{master3}) is negligible also in this case. In
both cases, the probability distribution $P_k(t_n)$ evolves as a spreading
gaussian. The long--term evolution of the entropy is markedly different for
these systems than for the periodic case. In Fig.~\ref{fig:entropy}, the time
evolution of the entropy $S$  for the QKR and the random kicked rotor are
compared. In the case of the QKR, the entropy stops growing when the constraint
is satisfied, while it grows without bound in the other case.

Finally, we remark that the approach of separating the Schr\"{o}dinger equation
into a master equation part supplemented by a term which takes into account
quantum interference effects,  provides a new insight on quantum diffusion and
dynamical localization. As we have shown, the localized or diffusive character
of the dynamics can be understood as the result of the competition between
these terms of the evolution equation. The Markovian term dominates the
dynamics for a time which depends on the topology of the dynamical response
spectrum. The general procedure has been illustrated  using the quantum kicked
rotor, in its periodic, quasi--periodic and random kick versions. In
particular, we have explained the exponentially localized probability
distribution of the QKR using simple, well accepted results from the theory of
stochastic processes. This exponential distribution results from the
constrained maximization of the coarse--grained entropy. The
localization--length constraint is due to the discrete nature of the dynamical
response spectrum of the QKR. In the other cases considered the dynamical
response spectrum is dense, the localization length is infinite ({\it i.e.} the
constraint is absent) and the entropy increase continues indefinitely, as is
characteristic of a diffusive process. We have shown, that in this case the
Schr\"{o}dinger equation is equivalent to a diffusion equation with the
classical diffusion coefficient.

The approach that we have presented here may be used to describe the dynamics
of an arbitrary quantum system in which there is quantum diffusion.

\emph{We acknowledge the support of PEDECIBA and CONICYT-Clemente Estable
(project \#6026), RD acknowledges partial financial support from\\
MCT/FINEP/CNPq (PRONEX) under contract 41.96.0886.00.}

\end{document}